# Link between spin fluctuations and Cooper pairing in copper oxide superconductors

K. Jin, N. P. Butch, K. Kirshenbaum, J. Paglione & R. L. Greene [*]

Center for Nanophysics & Advanced Materials and Department of Physics, University of Maryland, College Park, MD 20742, USA

**Although it is generally accepted that superconductivity (SC) is unconventional in the high-transition temperature copper oxides (high-$T_c$ cuprates), the relative importance of phenomena such as spin and charge (stripe) order, SC fluctuations, proximity to a Mott insulator, a pseudogap phase, and quantum criticality are still a matter of great debate[1]. In electron-doped cuprates, the absence of an anomalous pseudogap phase in the underdoped region of the phase diagram[2] and weaker electron correlations[3,4], suggest that Mott physics and other unidentified competing orders are less relevant and that antiferromagnetic (AFM) spin fluctuations are the dominant feature. Here we demonstrate that a linear-temperature ($T$-linear) scattering rate - a key feature of the anomalous normal state properties of the cuprates - is correlated with the Cooper pairing (SC). Through a study of magnetotransport in thin films of the electron-doped cuprate La$_{2-x}$Ce$_x$CuO$_4$ (LCCO), we show that an envelope of $T$-linear scattering surrounds the SC phase, and survives to zero temperature when superconductivity is suppressed by magnetic fields. Comparison with similar behavior found in organic superconductors[5] strongly suggests that the $T$-linear resistivity is caused by spin-fluctuation scattering. Our results establish a fundamental connection between AFM spin fluctuations and the pairing mechanism of high temperature superconductivity in the cuprates.**

*T*-linear resistivity is well known to appear in proximity to an AFM quantum critical point (QCP), as found in organic[5] and heavy-fermion[6] strongly correlated materials. Unlike the hole-doped cuprates, the absence of anomalous pseudogap physics and other unidentified competing phases in these materials allows such non-Fermi liquid (FL) properties to be attributed to the presence an AFM QCP (ref. 6).This has led to models which ascribe *T*-linear resistivity to a mechanism involving spin fluctuation scattering[7-9]. The case for this is particularly strong in the Bechgaard class of organic superconductors $(TMTSF)_2PF_6$, where *T*-linear scattering dominates the normal state transport above a SC state induced by the suppression of a spin-density wave (SDW) order by applied pressure[5]. The anisotropic two dimensional nature of the Bechgaard compounds allows for microscopic calculations of the interdependence of AFM and superconducting correlations[10], yielding a thorough understanding of the origin of the anomalous scattering rate in this case[5,11]. However, in general, no microscopic theory yet exists for the origin of *T*- linear scattering at low temperatures. In $(TMTSF)_2PF_6$, the *T*-linear scattering rate found at the SDW QCP has been shown to be suppressed with pressure along with the superconducting transition, with a scattering coefficient that approaches zero along with $T_c$ (refs. 5,11). In electron-doped $Pr_{2-x}Ce_xCuO_4$ (PCCO), *T*-linear resistivity is found down to 35 mK at *x* = 0.17 (ref. 12). Along with other evidence for a Fermi surface (FS) reconstruction[2,13-15], this observation suggests that an AFM QCP occurs near *x* = 0.17 in PCCO.

LCCO is an electron-doped cuprate[16] with properties very similar to PCCO, but with a SC dome that is slightly shifted toward lower Ce concentrations such that the SC phase exists for $0.06 \leq x \leq 0.17$ and is suppressed for *x* > 0.17. The phase diagram of LCCO (Fig. 1), constructed from our present transport studies on optimal to overdoped thin films (*x* ≥ 0.11) and prior work[17-19] for *x* < 0.12, has four distinct regions: the SC phase, the *T*-linear ($\rho \sim T$) region, the non-FL ($\rho \sim T^{1.6}$) region and the FL ($\rho \sim T^2$) region. In the SC doping range, all films exhibit a $T$-linear resistivity above $T_c$ that extends from $T_c$ up to a characteristic crossover temperature $T_1$, forming a shell of anomalous scattering that encases the SC dome. For example, the resistivity of optimally doped $x$ = 0.11 is linear from $T_c$ up to $T_1 \sim 45$ K (Supplementary Fig. 1). For higher doping, this temperature range (and thus $T_1$) decreases, tending toward zero along with $T_c$ itself at the end of the SC dome at a critical doping of $x_c = 0.175 \pm 0.005$. In PCCO, a similar phenomenon is observed (Supplementary Figs 2 and 3) with a $T$-linear region above $T_c$ that extends as far as films can be synthesized (*i.e.*, up to $x = 0.19$). Similarly, in hole-doped $La_{2-x}Sr_xCuO_4$ (LSCO) a $T$-linear component was shown to diminish upon approach of the end of the doping range of superconductivity[20], suggestive of a common relation between scattering and pairing in both electron- and hole-doped cuprates. The nature of the QCP in hole-doped cuprates remains uncertain. Note, however, that a linear resistivity identical to that of LSCO (ref. 20) was observed in $La_{1.6-x}Nd_{0.4}Sr_xCuO_4$ (Nd-LSCO) (ref. 21) at the QCP where stripe order is known to end.

A direct relation between $T$-linear scattering and $T_c$ is revealed through the doping dependence of each. As shown in Fig. 2, the scattering coefficient $A_1(x)$, obtained from fits to the $T$-linear regions with $\rho(T) = \rho_0 + A_1(x)T$, decreases with $T_c$ as $x$ is increased and approaches zero at the critical doping $x_c$. This scaling of $A_1$ with $T_c$ is also observed in PCCO (Supplementary Fig. 4), indicating that it is not specific to the doping concentration (which is shifted in PCCO as compared to LCCO for a given $T_c$), but is representative of a central relationship between the two energy scales ($T_c$ and $A_1$). The same relation has been found in $(TMTSF)_2PF_6$ (refs 5, 11), reflecting the intimate connection between the strength of the $T$-linear inelastic scattering and the electron pairing in systems governed by spin fluctuations. Similar scaling is seen in the hole-doped cuprates LSCO, Nd-LSCO, and $Tl_2Ba_2CuO_{6+\delta}$ (Tl2201) (ref. 11), again suggesting that the physics of scattering and pairing is the same in electron- and hole-doped cuprates.

The $T$-linear scattering is robust and survives in magnetic fields exceeding the SC upper critical field of LCCO. In fact, when superconductivity is completely suppressed, the $T$-linear resistivity extends down to the $T = 0$ limit without any indication of saturation or change in behavior. For instance, for $x = 0.15$ at 7.5 T (Fig. 3a, Supplementary Fig. 1), $T$-linear resistivity extends from $T \sim 20$ K down to the lowest measured temperature of 20 mK. Spanning over three decades in temperature, this behavior clearly points to a scattering mechanism that originates from an anomalous ground state. Similar behavior is found at higher $x$ (Fig. 3b), but occurs over a decreasing range as $T_c$ is suppressed to zero with doping, again suggesting that $T$-linear scattering is intimately tied to the presence of superconductivity.

Many experiments have shown that spin fluctuations dominate the physical properties in proximity to a critical doping under the SC dome in the more studied electron-doped cuprates PCCO and $Nd_{2-x}Ce_xCuO_4$ (NCCO) (refs 2, 22, 23). In analogy with these other electron-doped cuprates, it is expected that the boundary of AFM order in LCCO extrapolates to a QCP beneath the SC dome (indicated as $x_{FS}$ in Fig. 1), having a fundamental role in generating the SC phase. In particular, the extended $T$-linear transport scattering that persists to the lowest measurable temperatures is exactly in line with that expected at an AFM QCP for a two-dimensional disordered Fermi liquid system[9]. Moreover, inelastic neutron scattering experiments on electron-doped $Pr_{1-x}LaCe_xCuO_{4+\delta}$ show that the strength of the spin fluctuations decreases with overdoping in the SC phase and that these fluctuations disappear at the end of the SC dome[24].

Non-SC films of LCCO doped beyond $x_c$ exhibit a $T^2$ dependence of $\rho(T)$ in the low temperature limit, indicating a conventional FL behavior due to electron-electron scattering similar to that exhibited by $(TMTSF)_2PF_6$ (ref. 5) and LSCO (ref. 25). For example, LCCO films with $x = 0.18$ exhibit a $T^2$ resistivity up to 5 K, spanning over two orders of magnitude in temperature (Supplementary Fig. 5). The highest temperature of the quadratic behavior, $T_{FL}$, increases with increasing $x$ as shown for $x = 0.19$ and $0.21$ (Figs 3c and d), and notably, this line extrapolates to $T = 0$ at $x_c$.

In LCCO, the critical doping $x_c$ is exactly where the SC dome terminates and the two characteristic crossover temperatures $T_1$ and $T_{FL}$ approach absolute zero. Interestingly, indications of the singular nature of $x_c$ are evident even from within the overdoped FL regime of LCCO. In this region of the phase diagram, the coefficient of electron-electron scattering $A_2(x)$ (*i.e.*, obtained from fits to $\rho(T) = \rho_0 + A_2(x)T^2$) exhibits a strong enhancement upon approach to $x_c$ from higher doping, reminiscent of critical scattering upon approach to a QCP (ref. 6). This suggests that the onset of superconductivity marks a dramatic change in the ground state and its excitations. While FL behavior of resistivity has been reported at one doping in both hole-doped LSCO (ref. 25) and Tl2201 (ref. 26), such doping-tuned critical behavior in the non-SC region was not observed. In LCCO, the resistivity directly above the critical point at $x = 0.175$ and in the entire temperature regime above the characteristic temperatures $T_1$ and $T_{FL}$ is best fit by a single power law dependence, $\rho = \rho_0 + A'T^n$ with $n \approx 1.6$, up to at least 50 K (Supplementary Fig. 6). Perhaps not coincidentally, the same power law is observed above the FL ($\sim T^2$) regime in LSCO (refs 20, 25), signifying that scattering throughout the non-FL regime is governed by the same physics in both hole- and electron-doped cuprates. Clearly, our observation of critical behavior at $x_c$ will require further experimental and theoretical investigation to determine its significance for the unusual transport properties of the cuprates.

With the absence of anomalous pseudogap phenomena in electron-doped cuprates[2], comparisons to similarly tractable systems allow for far-reaching conclusions to be drawn. Studies[5,10,11] of the organic superconductor $(TMTSF)_2PF_6$ show that electron pairing and $T$-linear scattering arise from AFM (SDW) spin fluctuations. Given the very similar experimental transport properties and evolution of ground states in the phase diagram of LCCO, it is likely that the scattering and pairing in the electron-doped cuprates is governed by a similar interplay of spin fluctuations and superconductivity. The results of our work reported here, and their analogy to $(TMTSF)_2PF_6$ strongly suggests that the pairing in electron-doped cuprates is not coming from phonons or any other unusual pseudogap order parameter (such as $d$-density waves, orbital currents or stripe order), but rather from spin fluctuation-mediated pairing[27-29]. The striking similarities between transport properties of electron- and hole-doped cuprates provides evidence that the mechanism of the anomalous $T$-linear scattering rate and high-$T_c$ pairing are shared between the two families, and, furthermore, bear a striking resemblance to simpler systems well described by the spin fluctuation scenario. While the role of the pseudogap and unidentified competing phases in the hole-doped cuprates remains to be conclusively determined, the similar correlation between the $T$-linear scattering and $T_c$ for both electron-and hole-doped cuprates suggests that spin fluctuations also play the crucial role in hole-doped cuprates.

**Supplementary Information** is linked to the online version of the paper at www.nature.com/nature.

**Acknowledgements** We would like to thank L. Taillefer for extensive discussions and N. Doiron-Leyraud for some preliminary analysis of our zero-field data. We also appreciate important discussions with A. Chubukov, A. Millis, A.-M. S. Tremblay and C. Varma. Some experimental help was provided by X. Zhang, P. Bach and G. Droulers. This research was supported by the NSF under DMR-0952716 (J.P. and K.K.) and DMR-0653535 (R.L.G) and the Maryland Center for Nanophysics and Advanced Materials (K.J. and N.P.B.).

**Author Contributions** K.J. prepared and characterized the thin-film samples. K.J., N.P.B, K.K., and J.P. performed the transport measurements and data analysis. N.P.B., J.P, and R.L.G wrote the manuscript. R.L.G. conceived and directed the project.

**Author Information** Reprints and permissions information is available at www.nature.com/reprints. The authors declare no competing financial interests. Readers are welcome to comment on the online version of this article at www.nature.com/nature. Correspondence and requests for materials should be addressed to R.L.G. (rgreene@squid.umd.edu).

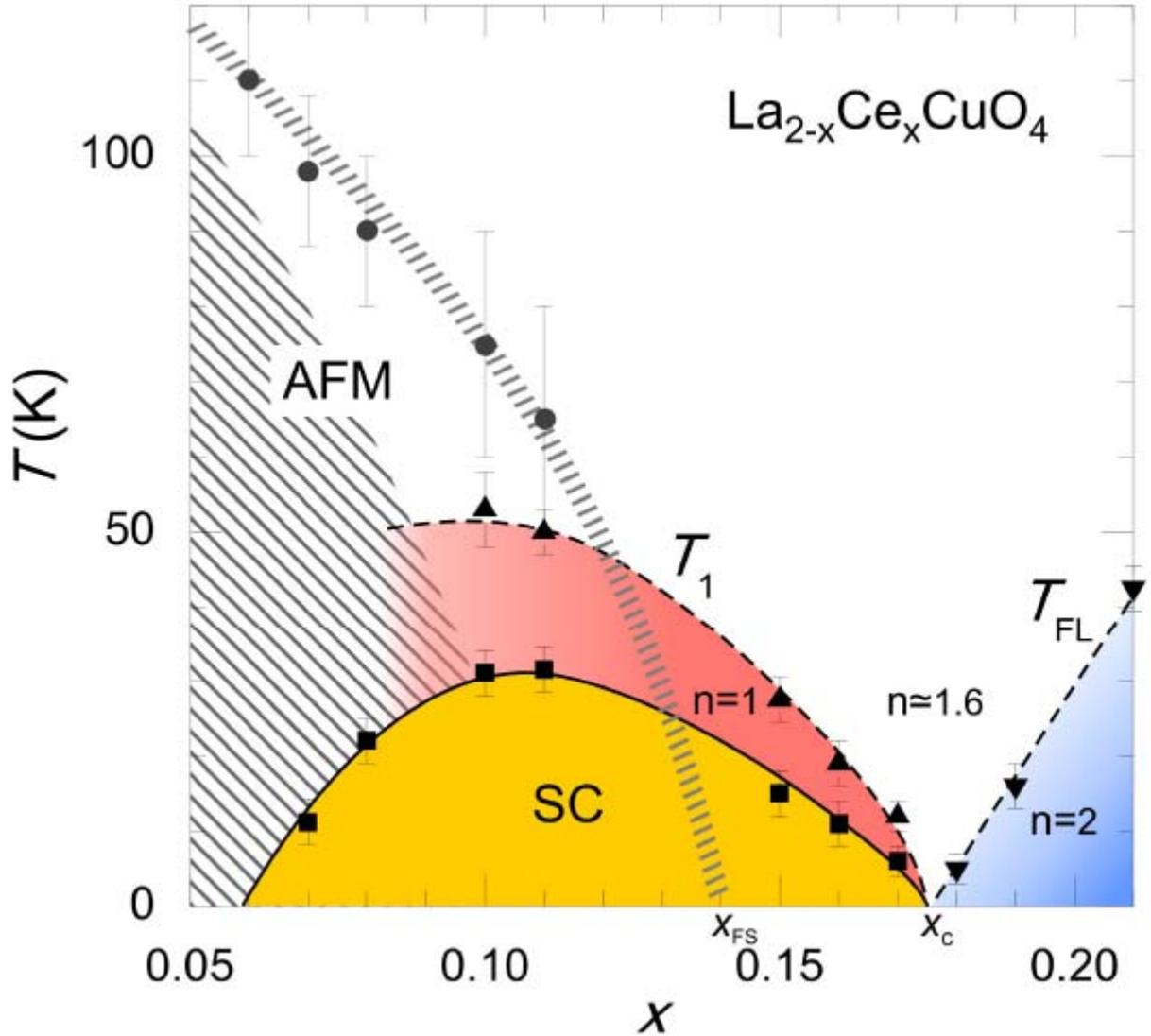

**Figure 1 | Temperature-doping ($T$, $x$) phase diagram of $La_{2-x}Ce_xCuO_4$ (LCCO).** The resistivity in zero field can be expressed by $\rho = \rho_0 + AT^n$, with n = 1 and 2 for the red and blue regimes, respectively. Between the $\rho \sim T$ (n = 1) and the Fermi liquid (n = 2) regimes, the data below 50 K could be well fit by a single power low with n ≈ 1.6. The yellow regime is the superconducting (SC) dome. The SC, $\rho \sim T$ and Fermi liquid regimes terminate at one critical doping, $x_c$. The temperatures $T_1$ (triangles) and $T_{FL}$ (inverted triangles) mark the crossover temperatures to the $\rho \sim T$ and Fermi liquid regimes, respectively. In order to illustrate the $\rho \sim T$ regime more clearly, the boundary of the SC dome (squares) is defined as the lowest temperature

of the linear-in-$T$ resistivity for $x \geq 0.1$. For $x < 0.1$, the resistivity shows an upturn (hatched area) with decreasing temperature, a typical feature for underdoped cuprates. Owing to the upturn, the SC boundary for $x < 0.1$ is defined as the temperature where the resistivity reaches zero ($T_{c0}$). The AFM (or spin density wave, SDW) regime (circles) is estimated from previous in-plane angular magnetoresistance (AMR) measurements[18]. A quantum critical point (QCP) associated with a SDW Fermi surface reconstruction is estimated to occur near $x = 0.14$ (indicated as $x_{FS}$). LCCO can only be prepared in thin film form and, therefore, the evidence for a SDW (antiferromagnetic) QCP under the SC dome is not as conclusive as for the electron-doped cuprates $Pr_{2-x}Ce_xCuO_4$ or $Nd_{2-x}Ce_xCuO_4$. Nevertheless, in LCCO the change of the sign of the low $T$ Hall coefficient at $x \sim 0.14$ (ref. 19), AMR data, and a low temperature metal to insulator crossover at $x \sim 0.14$ (ref. 16) all suggest that such a QCP, associated with FS reconstruction, does occur near $x = 0.14$. The error bars on the circles are from ref. 18 and those on other symbols represent the standard error in the fit to the data.

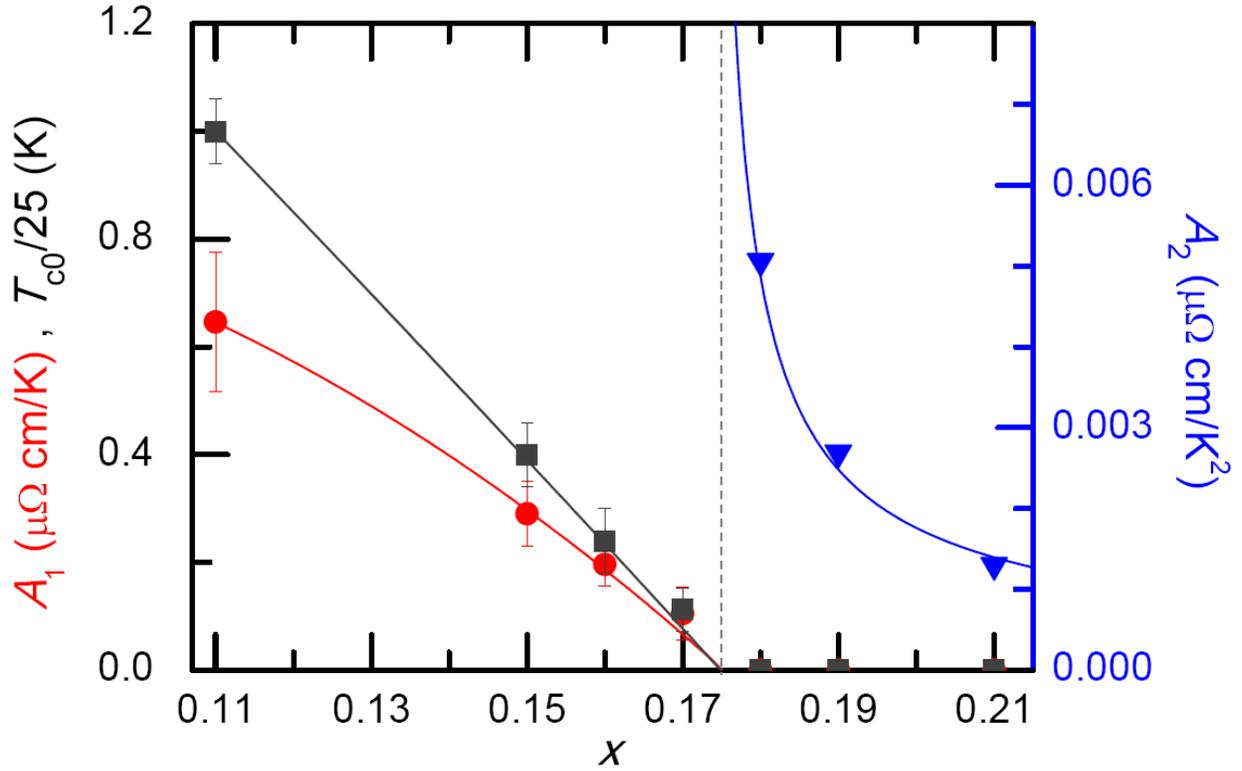

**Figure 2 | Doping dependence of scattering rates, $A_1$ and $A_2$, in zero field.** Left axis: $A_1$ (red circles) and also $T_{c0}$ (divided by 25, black squares) versus $x$. Right axis: $A_2$ (blue triangles) versus $x$. For the superconducting LCCO films with $x < 0.18$, $A_1$ data are obtained from the $\rho \sim T$ region ($\rho = \rho_0 + A_1 T$, red regime in Fig. 1). The error bars are the standard deviation over many samples of each doping. Note that in the optimally doped region, the highest superconducting transition temperatures of $x = 0.1$ and $0.11$ are almost the same by a slight oxygen variation, and their resistivity also shows similar behavior. Thus, only one nominal $x = 0.1$ sample was studied here, nevertheless, both the $A_1$ and $T_{c0}$ (data not shown) fall into the statistical error of the $x = 0.11$ samples. We use the $x = 0.11$ doping to represent the optimal doping level here. For the non-superconducting films with $x \geq 0.18$, $A_2$ data are obtained from the $\rho \sim T^2$ region ($\rho = \rho_0 + A_2 T^2$, blue regime in Fig. 1). It is noteworthy that the amplitude of the $T$-linear scattering scales with the SC transition temperature (both ending around $x_c = 0.175$), reflecting the intimate

relation between the *T*-linear scattering rate and the superconductivity. From the non-SC side, as the doping approaches $x_c$ from higher doping, the coefficient of electron-electron scattering increases very quickly, reminiscent of critical scattering upon approach to a quantum critical point.

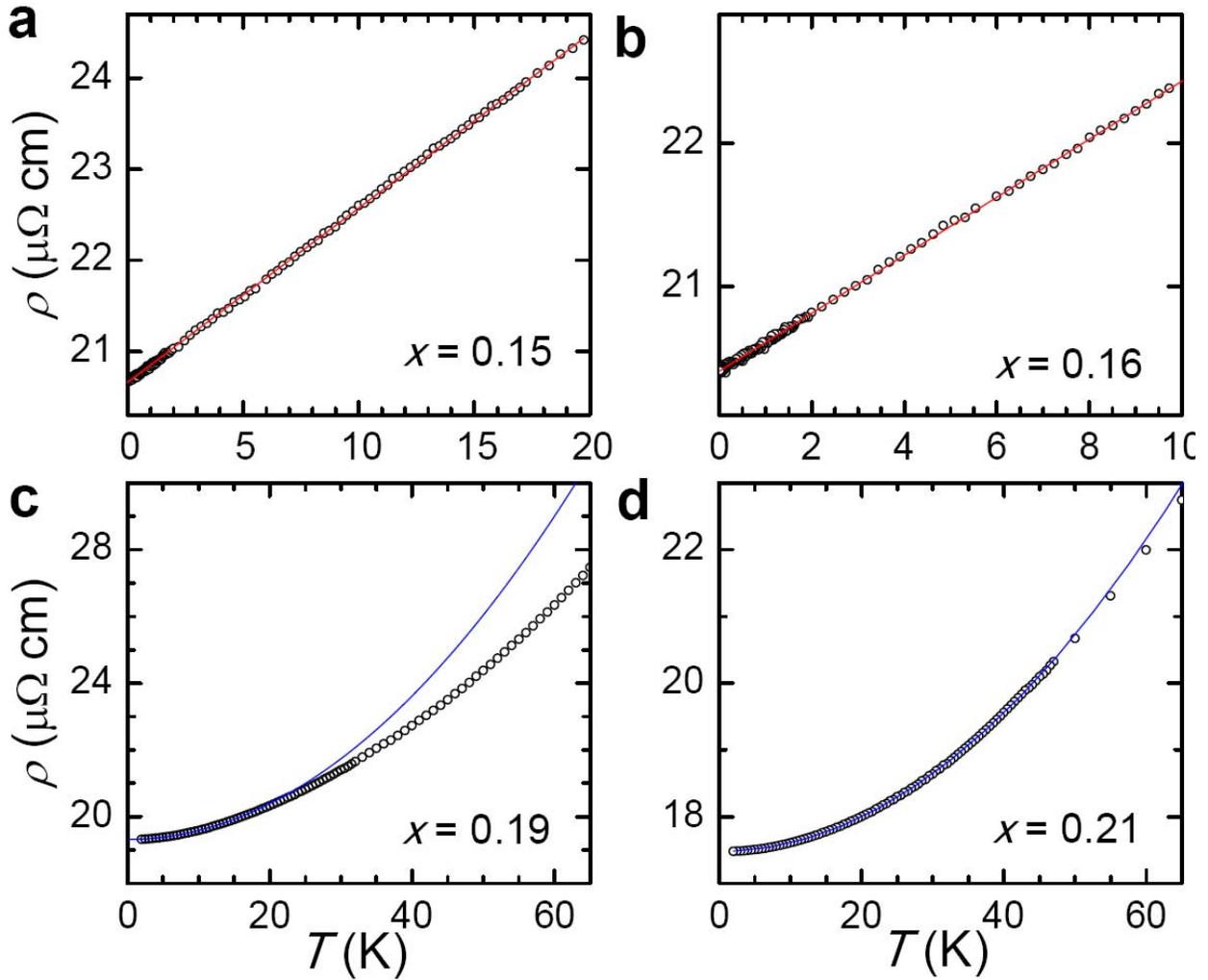

**Figure 3| Temperature dependence of normal-state resistivity. a** and **b**, $\rho(T)$ of $x$ = 0.15 and 0.16 LCCO films in a perpendicular magnetic field where the superconductivity is just suppressed, i.e., at 7.5 and 7 T, respectively. The data can be fitted by $\rho = \rho_0 + A_1 T$ down to the lowest measuring temperature. The linearity of the resistivity of $x$ = 0.15 persists from 20 K

down to 20 mK, spanning over three decades in temperature. That is, the $\rho \sim T$ region shown in Fig. 1 can extend down to the $T = 0$ limit, pointing to a scattering mechanism that originates from an anomalous ground state. **c** and **d,** $\rho(T)$ of $x = 0.19$ and $0.21$ in zero field, fitted by $\rho = \rho_0 + A_2 T^2$ (blue lines). In the non-SC regime ($x \geq 0.18$), the Fermi liquid behavior can also persist to the lowest temperature, i.e., down to 20 mK (as seen in supplementary Fig. 5 for $x = 0.18$).

## Supplementary Information

**Samples.** The c-axis-oriented $La_{2-x}Ce_xCuO_4$ (LCCO) and $Pr_{2-x}Ce_xCuO_4$ (PCCO) films were deposited directly on (100) $SrTiO_3$ substrates by a pulsed laser deposition (PLD) technique utilizing a KrF excimer laser as the exciting light source[18]. The films were typically 100-150 nm in thickness. Since the oxygen content has an influence on both the SC and normal state properties of the material[13], we took extra care in optimizing the annealing process for each Ce concentration. We prepared many films with variable oxygen content, and found that the optimized samples showed metallic behavior down to the lowest measured temperature (20 mK), while non-optimized samples often show an upturn (either in fields or at zero field) at low $T$, due to oxygen-induced disorders. Moreover, the optimized samples showed a narrow transition width (for superconducting samples) and low residual resistivity (for nonsuperconducting samples). Using these criteria, we found the best film growth conditions and excellent reproducibility of the transport data presented here. The films were patterned into Hall bar bridges using photolithography and ion milling techniques for the transport measurements.

**Measurements.** The high-$T$ measurements (above 2 K) were carried out in a 14T PPMS and low-$T$ measurements were done in an Oxford dilution fridge (down to 20 mK), with an overlapped temperature range. For example, the linear-in-$T$ resistivity data shown in Fig. 3a were measured in the dilution fridge from 20 mK to 20 K. Electrical current was applied in the ab-plane while the magnetic field was applied along the c-axis.

**The boundary of the SC dome.** The boundaries of SC for both LCCO (Fig. 1) and PCCO (Fig. S3) are determined as the temperature at which the resistivity starts to deviate from the linear-in-$T$ resistivity behavior as seen in Figures S1 and S2, respectively. For the underdoped samples, since there is an upturn of the resistivity above the superconducting transition, we define the zero resistive state temperature ($T_{c0}$) as the superconducting boundary.

**Universal relation between $A_1$ and $T_c$ for PCCO and LCCO.** A comparison between PCCO and LCCO shows that the $A_1$ to $T_c$ scaling is a universal behavior in electron-doped cuprates. Because of the different optimal doping levels, we normalize the $A_1(x)$ to the coefficient at the optimal doping, i.e., $A_{opt} = A_1(x = 0.15)$ for PCCO and $= A_1(x = 0.11)$ for LCCO. The x axes are plotted with the same interval ($\Delta x = 0.01$) but with different starting points ($x = 0.15$ for PCCO and $= 0.11$ for LCCO) for the comparison shown in Fig. S4. The doping dependence of the normalized coefficient data ($A_1(x)/A_{opt}$) of PCCO and LCCO fall onto one straight line, suggesting the linear resistivity in PCCO would disappear at a doping $x \sim 0.215$. The normalized superconducting transition temperature ($T_c(x)/T_c^{opt}$) also falls onto one straight line. That is, the superconducting transition temperature scales with the coefficient of the linear term, reflecting the intimate relation between the strength of the linear resistivity and the pairing. We note that the $A_1(x)/A_{opt}$ scales monotonically with $T_c(x)/T_c^{opt}$ in the overdoped region, but a maximum is found between $x = 0.16$ and $0.17$ for PCCO (see Fig. S4b). This change is most likely associated with the Fermi surface reconstruction quantum critical point found at this doping[13-15]. Below this QCP, the low $T$ resistivity starts to have an upturn, which also adversely impacts the coefficient of any attempted linear fit. This is also true for LCCO (ref. 16). In LCCO, the normalized $A_1(x)$ values suggest that an analogous QCP would exist between $x = 0.12$ and $x = 0.14$, which is consistent with the extrapolated AFM/SDW endpoint in the phase diagram (Fig. 1).

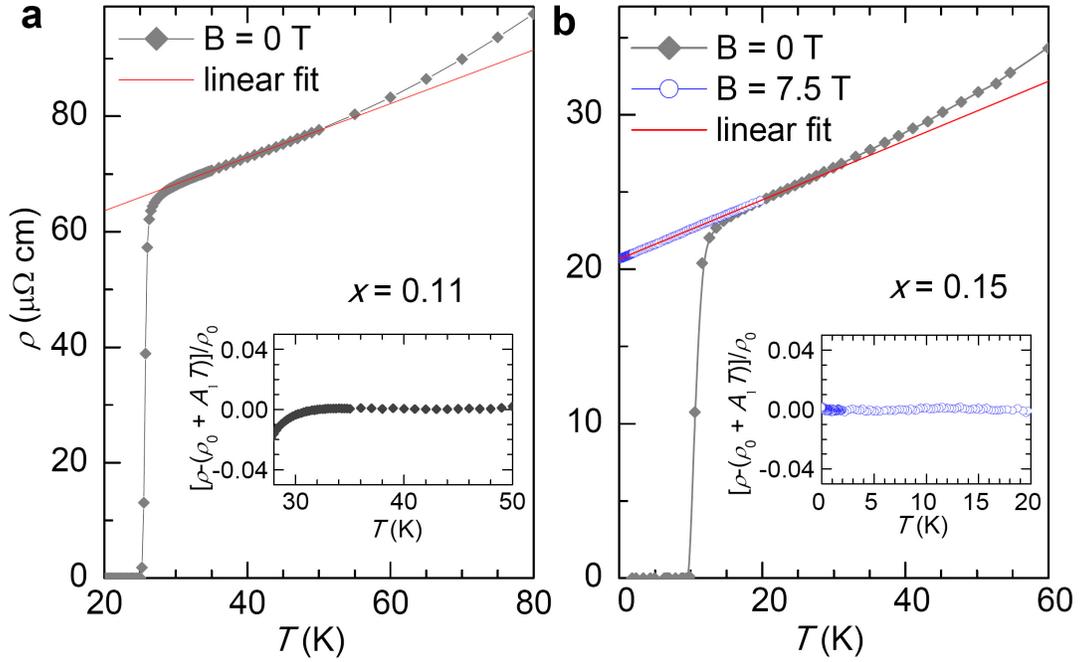

**Figure S1 | Temperature dependence of resistivity in superconducting La$_{2-x}$Ce$_x$CuO$_4$ (LCCO). a,** $\rho$ (T) of optimally doped LCCO with $x$ = 0.11 in zero field (diamonds), fitted by $\rho$ (T) = $\rho_0$ + $A_1 T$ (red line). **b,** $\rho$ (T) of $x$ = 0.15 at 0 (diamonds) and 7.5 T (circles). The red line is the linear fit to the 7.5 T data. The insets show the fitting quality presented as $\Delta \rho / \rho$ vs. T, where $\Delta \rho = \rho - (\rho_0 + A_1 T)$.

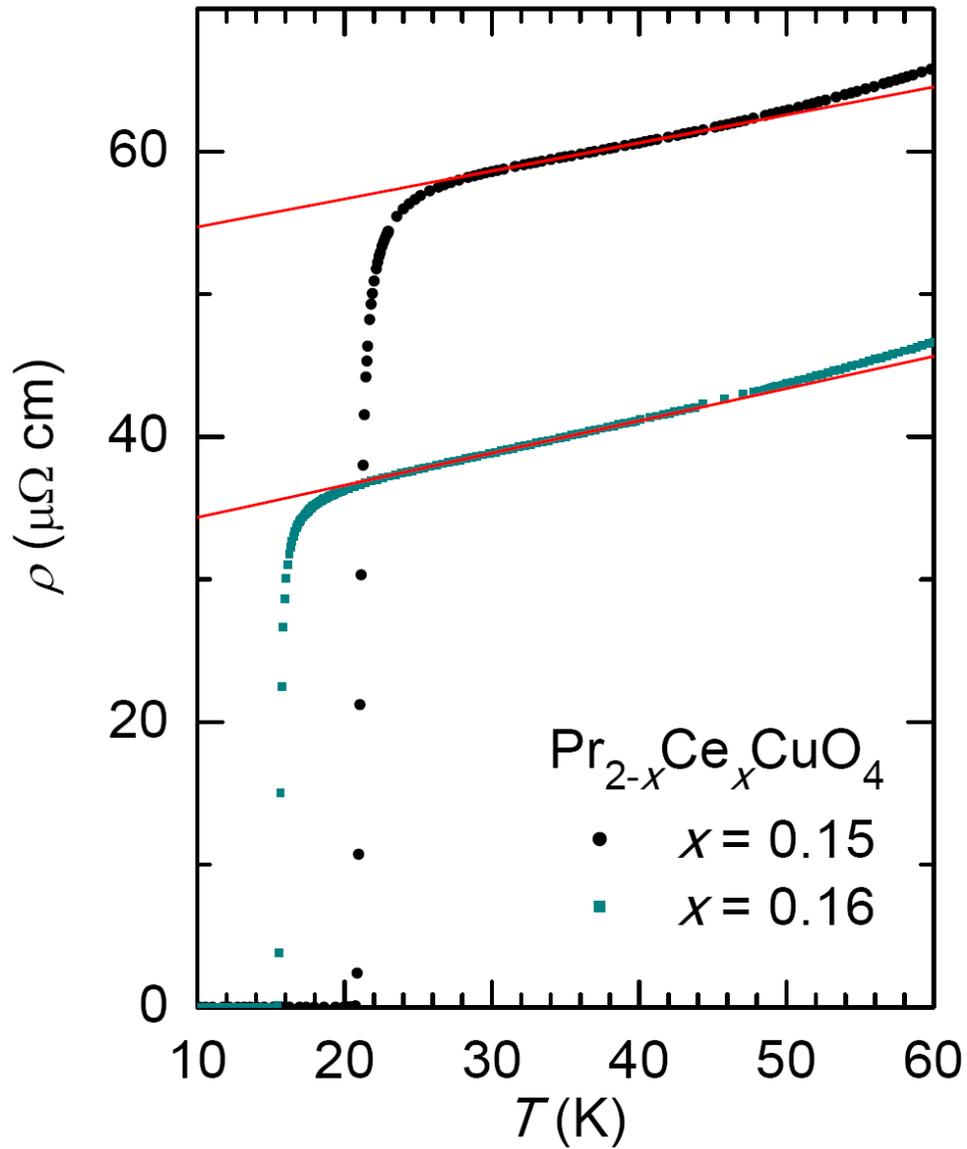

**Figure S2 | Linear-in-T dependence of resistivity in superconducting Pr$_{2-x}$Ce$_x$CuO$_4$ (PCCO) films.** $\rho$ (T) of PCCO with $x$ = 0.15 and 0.16 in zero field, fitted by $\rho(T) = \rho_0 + A_1 T$ (red lines). The data are from ref. 13.

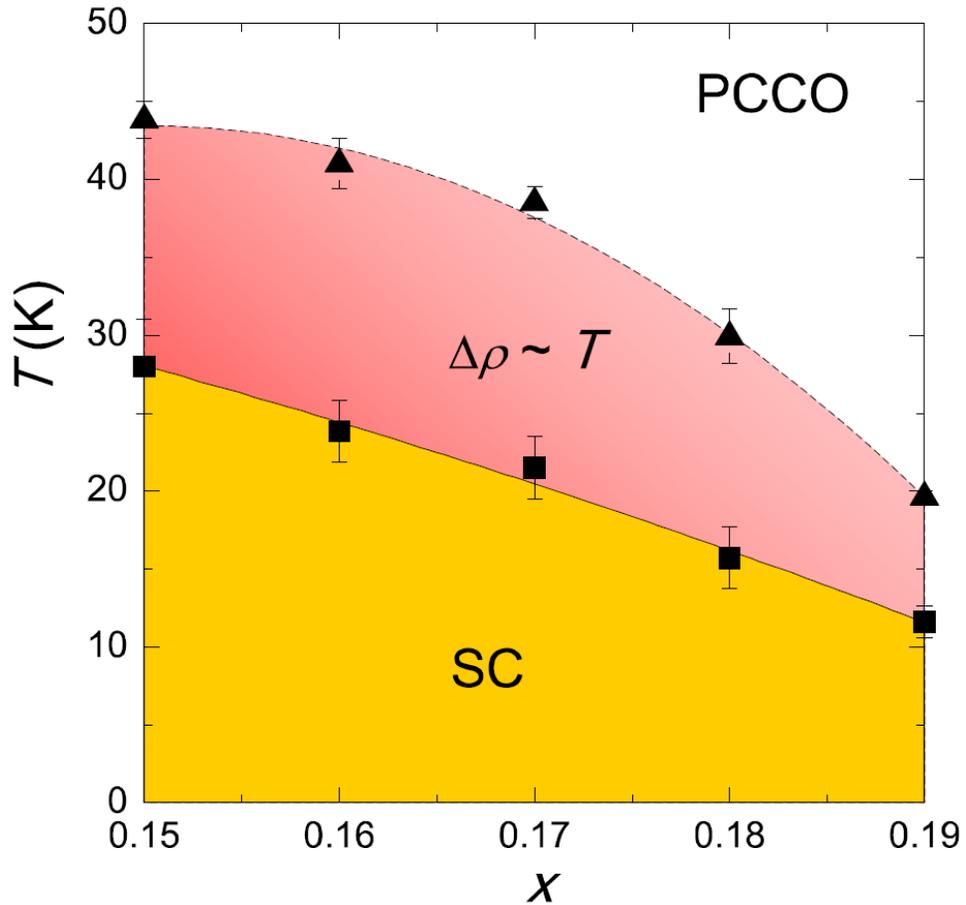

**Figure S3 | Temperature-doping (*T*, *x*) phase diagram of PCCO in zero field.** The phase diagram of PCCO with 0.15 ≤ x ≤ 0.19 shows superconducting (SC) and *T*-linear regions. The boundaries of SC and *T*-linear regions are defined as the starting and ending temperatures of the *T*-linear resistivity (see Fig. S2). Error bars represent the standard error in the fit.

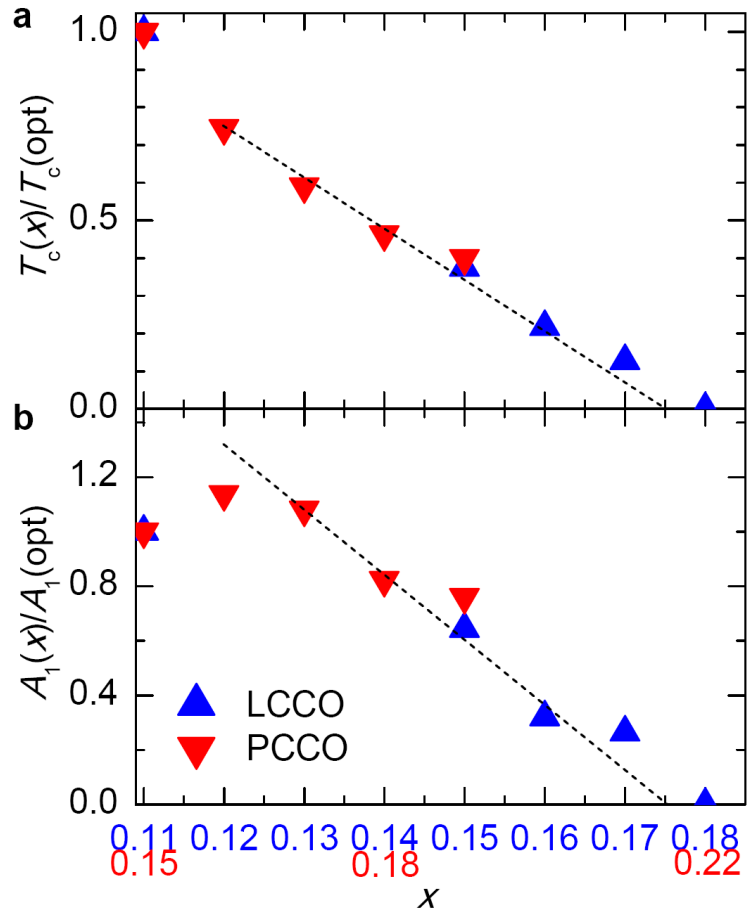

**Figure S4 | Relation between the superconducting transition temperature and the scattering rate in LCCO and PCCO. a,** Doping dependence of reduced $T_c$, is normalized to the superconducting transition temperature of the optimal doping at $x = 0.11$ and $0.15$ for LCCO and PCCO, respectively. **b,** Doping dependence of reduced $A_1(x)$ (normalized to that of optimal doping), the coefficient of the linear resistivity in the $\rho \sim T$ region. The x axes for LCCO and PCCO start from their optimal doping levels, $x = 0.11$ and $0.15$, respectively (see text for details).

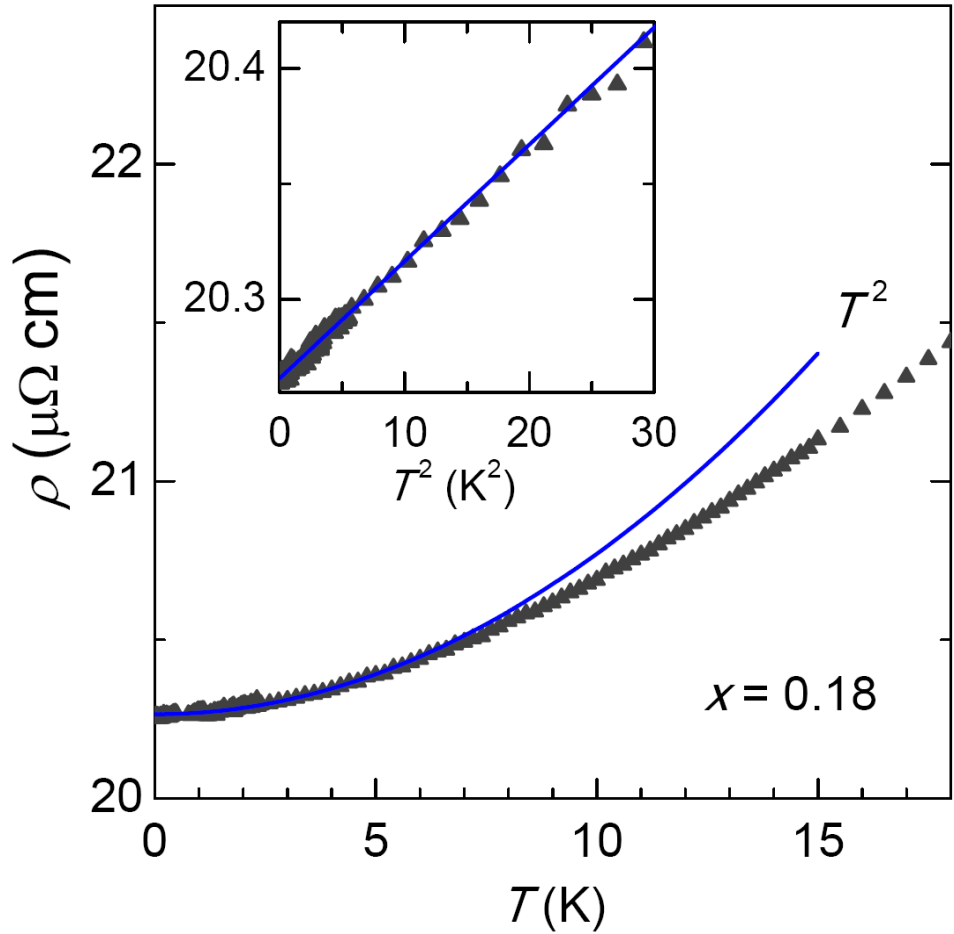

**Figure S5 | Temperature dependence of resistivity in LCCO with x = 0.18.** $\rho(T)$ of $x = 0.18$ in zero field (triangles), fitted by $\rho(T) = \rho_0 + A_2 T^2$ (blue line). The resistivity shows a Fermi liquid (FL) behavior from 20 mK to ~5 K. Inset: $\rho(T)$ versus $T^2$ in the FL range.

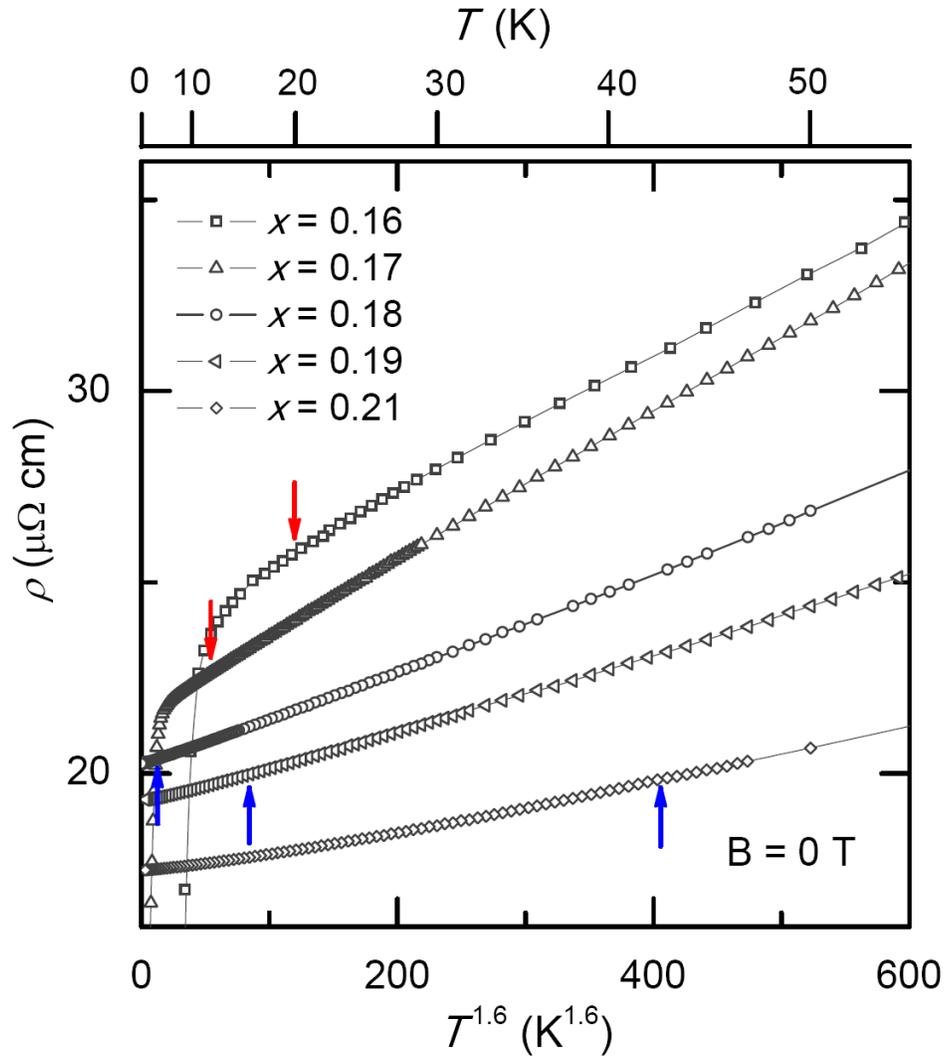

**Figure S6 | Temperature dependence of resistivity in LCCO from x = 0.16 to 0.21.** $\rho$ (T) of $x$ = 0.16, 0.17, 0.18, 0.19, and 0.21 in zero field (symbols), plotted against $T^{1.6}$. All these doping levels show an approximate behavior, $\rho \sim T^{1.6}$ somewhat above the crossover temperatures, $T_1$ (red arrows) and $T_{FL}$ (blue arrows). For clarity, we also put a linear scale of temperature on the top.